\documentclass[aps,prb,twocolumn,superscriptaddress,amsmath,amssymb]{revtex4}

\usepackage{amssymb}
\usepackage{graphicx}
\usepackage{bm}
\usepackage{epsfig}
\usepackage[ansinew]{inputenc}

\begin{document}

\title{Training-induced inversion of spontaneous exchange bias field on La$_{1.5}$Ca$_{0.5}$CoMnO$_{6}$}

\author{L. Bufai\c{c}al}
\email{lbufaical@if.ufg.br}
\affiliation{Instituto de F\'{\i}sica, Universidade Federal de Goi\'{a}s, 74001-970, Goi\^{a}nia, GO, Brazil}

\author{R. Finkler}
\affiliation{Instituto de F\'{\i}sica, Universidade Federal de Goi\'{a}s, 74001-970, Goi\^{a}nia, GO, Brazil}

\author{L. T. Coutrim}
\affiliation{Instituto de F\'{\i}sica, Universidade Federal de Goi\'{a}s, 74001-970, Goi\^{a}nia, GO, Brazil}

\author{P. G. Pagliuso}
\affiliation{Instituto de F\'{\i}sica ``Gleb Wataghin", UNICAMP, 13083-859, Campinas, SP, Brazil}

\author{C. Grossi}
\affiliation{Centro Brasileiro de Pesquisas F\'{\i}sicas, 22290-180, Rio de Janeiro, RJ, Brazil}

\author{F. Stavale}
\affiliation{Centro Brasileiro de Pesquisas F\'{\i}sicas, 22290-180, Rio de Janeiro, RJ, Brazil}

\author{E. Baggio-Saitovitch}
\affiliation{Centro Brasileiro de Pesquisas F\'{\i}sicas, 22290-180, Rio de Janeiro, RJ, Brazil}

\author{E. M. Bittar}
\affiliation{Centro Brasileiro de Pesquisas F\'{\i}sicas, 22290-180, Rio de Janeiro, RJ, Brazil}

\date{\today}

\begin{abstract}
In this work we report the synthesis and structural, electronic and magnetic properties of La$_{1.5}$Ca$_{0.5}$CoMnO$_{6}$ double-perovskite. This is a re-entrant spin cluster material which exhibits a non-negligible negative exchange bias effect when it is cooled in zero magnetic field from an unmagnetized state down to low temperature. X-ray powder diffraction, X-ray photoelectron spectroscopy and magnetometry results indicate mixed valence state at Co site, leading to competing magnetic phases and uncompensated spins at the magnetic interfaces. We compare the results for this Ca-doped material with those reported for the resemblant compound La$_{1.5}$Sr$_{0.5}$CoMnO$_{6}$, and discuss the much smaller spontaneous exchange bias effect observed for the former in terms of its structural and magnetic particularities. For La$_{1.5}$Ca$_{0.5}$CoMnO$_{6}$, when successive magnetization loops are carried, the spontaneous exchange bias field inverts its sign from negative to positive from the first to the second measurement. We discuss this behavior based on the disorder at the magnetic interfaces, related to the presence of a glassy phase. This compound also exhibits a large conventional exchange bias, for which there is no sign inversion of the exchange bias field for consecutive cycles.
\end{abstract}

\maketitle

\section{INTRODUCTION}

The exchange bias (EB) effect is a well known phenomena characterized by a shift of the magnetization as a function of applied field ($H$) hysteresis loop along the $H$ axis. It is usually ascribed to the unidirectional exchange anisotropy at the interface between antiferromagnetic (AFM) and ferromagnetic (FM)/ferrimagnetic (FIM) phases in heterogeneous systems \cite{Nogues,Nogues2}, but it has also been observed at the interfaces of spin glass (SG) and AFM/FM/FIM phases \cite{Ali,Gruyters}. This effect is conventionally observed after cooling the system through its N\'{e}el temperature ($T$) in the presence of $H$ or if $H$ is applied during the materials's fabrication. Recently, improvements on oxide heterostructures growth have renewed the interest in such compounds, since it opens the real possibility of fabrication of multifunctional devices using strongly correlated electronic materials \cite{Dagotto}.

Recently, some compounds have been reported to exhibit a spontaneous EB effect, i.e., there is a shift along the field axis on the magnetization as a function of magnetic field measurement [$M(H)$] even when the system is cooled from an unmagnetized state and without the presence of external magnetic field \cite{Wang,Nayak,Maity,Murthy,Meu}. Despite the different mechanisms claimed to explain this zero field cooled exchange bias (ZEB) effect on distinct materials, two ingredients emerge as responsible for the phenomena: the presence of a glassy magnetic state re-entrant to a conventional magnetism and the reconfiguration of the spins during the initial field application at the $M(H)$ measurement.

The transition-metal (TM) oxides are potential candidates to exhibit the ZEB effect due to their intrinsic structural/magnetic inhomogeneity. From this class of materials, the perovskite-type (ABO$_{3}$, A-rare/alkaline-earth and B-TM) stands out. Ionic substitution at the A-site with cations of different charge and/or radii is a straightforward method to obtain interesting properties as glassy magnetic behavior \cite{Murthy2}, structural/magnetic phase separation \cite{Narayanan}, ferroelectricity \cite{Fukushima} and also conventional and spontaneous exchange bias effects \cite{Murthy,Meu}. The La$_{2-x}$Sr$_{x}$CoMnO$_{6}$ double-perovskite (DP) series exhibits the largest ZEB effect reported so far, for which 25\% of Sr substitution at La site leads to the largest shift on the $M(H)$ curve \cite{Murthy2}. Sr$^{2+}$ substitution at La$^{3+}$ site on the parent compound La$_{2}$CoMnO$_{6}$ leads to anti-site disorder, mixed valence of Co ions and competition between different magnetic phases, resulting in a re-entrant spin-glass (RSG) behavior \cite{Murthy}.

The microscopic mechanisms responsible for the ZEB effect observed on DP compounds are not yet understood, and in order to verify the currently proposed ideas it is important to study novel materials. In this context, two factors make La$_{1.5}$Ca$_{0.5}$CoMnO$_{6}$ a very interesting focus of investigation. Firstly, there is an advantage in doping La$_{2}$CoMnO$_{6}$ with Ca instead of Sr, which is the fact Ca$^{2+}$ ionic radius is closer to La$^{3+}$ than Sr$^{2+}$ \cite{Shannon}. As it is well known, DP oxides exhibit a strong correlation between their structural, electronic and magnetic properties. As such, it is always very difficult to predict which of these properties would be more strongly affected by doping, and consequently, would lead the variation of the others. The use of Ca, instead of Sr, on La$_{1.5}$A$_{0.5}$CoMnO$_{6}$ leads to presumable less structural changes in relation to La$_{2}$CoMnO$_{6}$, making it easier to notice the influence of electronic changes due ionic substitution. Moreover, the comparison between the Ca- and Sr-doped compounds provides insights about the influence of crystal structure to the ZEB effect.

In this work, we report the synthesis and investigation of structural, electronic and magnetic properties of the new DP compound La$_{1.5}$Ca$_{0.5}$CoMnO$_{6}$. It presents a cluster spin glass (CG) phase concomitantly with AFM and FM phases, being thus a RSG material. It also exhibits both the spontaneous and conventional exchange bias (CEB) effects. Our results are compared to those reported for the analogue compound La$_{1.5}$Sr$_{0.5}$CoMnO$_{6}$, and we discuss the differences in terms of structural and magnetic particularities of each compound.  Interestingly, when consecutive $M(H)$ loops are measured  on La$_{1.5}$Ca$_{0.5}$CoMnO$_{6}$ after cooling the system in the absence of applied magnetic field there is a sign inversion of the ZEB field from negative to positive values from the first to the second loop. We discuss this result on the basis of two contributions to the EB effect, the rotatable and frozen spins at the AFM/FM/CG interfaces.

\section{EXPERIMENT DETAILS}

Polycrystalline La$_{1.5}$Ca$_{0.5}$CoMnO$_{6}$ was synthesized by conventional solid-state reaction method. Stoichiometric amounts of La$_{2}$O$_{3}$, CaO, Co$_{3}$O$_{4}$ and MnO were mixed and heated at $800^{\circ}$C for 12 hours in air atmosphere. Later the sample was re-grinded before a second step of 24 hours at $1200^{\circ}$C. Finally the material was grinded, pressed into pellets and heated at $1300^{\circ}$C for 24 hours. High resolution x-ray powder diffraction (XRD) data was collected at room $T$ using a Bruker $D$8 $Discover$ diffractometer, operating with Cu K$_{\alpha}$ radiation over the angular range $10\leq\theta\leq100^{\circ}$, with a 2${\theta}$ step size of 0.01$^{\circ}$. Rietveld refinement was performed with GSAS software and its graphical interface program \cite{GSAS}. X-ray photoelectron spectroscopy (XPS) experiments were performed using an ultra-high vacuum (UHV) chamber equipped with a SPECS analyzer PHOIBOS 150. AC and DC magnetization measurements were carried out using a commercial SQUID-VSM magnetometer. DC data measurements were performed in both zero field cooled (ZFC) and field cooled (FC) modes. AC magnetization measurements were performed in oscillating field with amplitude $H_{ac}$ = 5 Oe, in six different frequencies in the range 25-10000 Hz.

\section{RESULTS AND DISCUSSION}

\subsection{Crystal structure}

\begin{figure}
\begin{center}
\includegraphics[width=0.5 \textwidth]{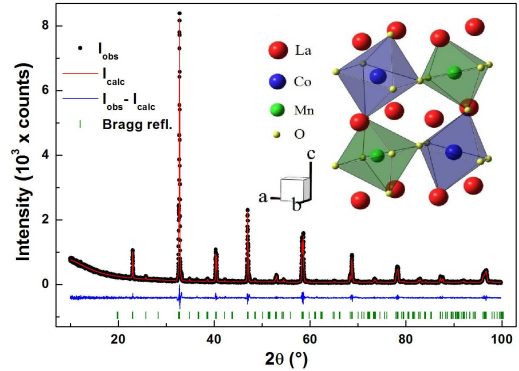}
\end{center}
\caption{Rietveld refinement fitting of La$_{1.5}$Ca$_{0.5}$CoMnO$_{6}$. The vertical lines represent the Bragg reflections and the upper panel shows the schematic structure of the DP.}
\label{FigXRD}
\end{figure}

La$_{1.5}$Ca$_{0.5}$CoMnO$_{6}$ room $T$ XRD data revealed the formation of a single phase DP (Fig. \ref{FigXRD}). Its structure could be successful refined in the monoclinic $P2_1/n$ space group, the same observed for the relative compound La$_{2}$CoMnO$_{6}$ \cite{Murthy2}. This is not surprising, since Ca$^{2+}$ and La$^{3+}$ ionic radii are very close in XII coordination (1.34 $\AA$ and 1.36 $\AA$ respectively) \cite{Shannon}.

The main parameters obtained from the Rietveld refinement are displayed in Table \ref{T1}. Comparing the average Co--O  (1.98 \AA) and Mn--O (1.95 \AA) bond lengths to those reported for La$_{2}$CoMnO$_{6}$ (2.04 $\AA$ and 1.94 $\AA$ respectively) \cite{Murthy2}, it can be noted that Co--O shrinks in relation to the undoped material while Mn--O remains nearly unchanged. The TM valences for La$_{2}$CoMnO$_{6}$ are believed to be Co$^{2+}$ and Mn$^{4+}$, thus the results suggest that Ca$^{2+}$ substitution at the La$^{3+}$ site acts mainly to change the Co valence. This is corroborated by XPS measurements, which indicate mixed valence Co$^{2+}$/Co$^{3+}$ (see below).

\begin{table*}
\centering
\caption{Structural parameters for La$_{1.5}$Ca$_{0.5}$CoMnO$_{6}$ obtained from Rietveld refinement.\label{T1}}
\begin{tabular}{ccccccccccccccccccc}
\hline
\hline
$a$ (\AA)    & $b$ (\AA)   & $c$ (\AA)    & $\beta$ ($^{\circ}$)   & Co--O (\AA) & Mn--O (\AA) & Co--O--Mn ($^{\circ}$) & $\chi^{2}$  & $R_{wp}$      \\
5.4843(1)  & 5.4498(1)  & 7.7099(2) & 89.94(1) & 1.981(14) & 1.954(14)& 158.8(5) & 1.7 & 9.6  \\
\hline
\hline
\end{tabular}
\end{table*}

Another important result observed is that Co--O--Mn bond angle for our Ca-doped sample ($158.8^{\circ}$) is smaller than those observed for the Sr-doped compounds La$_{1.75}$Sr$_{0.25}$CoMnO$_{6}$ ($163.8^{\circ}$) \cite{Murthy3}, La$_{1.6}$Sr$_{0.4}$CoMnO$_{6}$ ($165.3^{\circ}$) \cite{Gamari} and LaSrCoMnO$_{6}$ ($180^{\circ}$) \cite{Androulakis}, which is consistent to the fact that larger A-site ions lead to more symmetric crystal structure for DP compounds \cite{ReviewDP,Raveau}. Since the magnetic couplings between TM ions are strongly correlated to the bond angles, this result is probably one of the factors responsible for the differences between the magnetic properties of La$_{1.5}$Ca$_{0.5}$CoMnO$_{6}$ and La$_{1.5}$Sr$_{0.5}$CoMnO$_{6}$, as will be discussed next.

\subsection{X-ray photoelectron spectroscopy}

All the XPS results here displayed correspond to the use of monochromatic Al-$K_{\alpha}$ x-ray radiation ($h\nu=1486.6$ eV) using spectrometer pass energy ($E_{pass}$) of 15 eV. The spectrometer was previously calibrated using the Au 4$f_{7/2}$ (84.0 eV) which results on a full-width-half-maximum (FWHM) better than 0.7 eV, for a sputtered metallic gold foil. Prior to mounting the sample in UHV, it was slightly polished and ultra-sonicated sequentially in isopropyl alcohol and water. In the following, the sample was mounted on the top of a pure gold foil and referenced by setting the adventitious carbon C 1$s$ peak to 284.6 eV.

\begin{figure}
\begin{center}
\includegraphics[width=0.5 \textwidth]{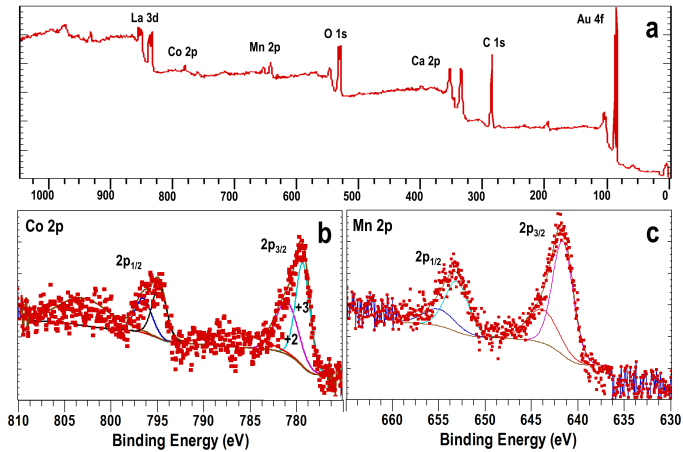}
\vspace{-0.8 cm}
\end{center}
\caption{(a) XPS survey spectra for La$_{1.5}$Ca$_{0.5}$CoMnO$_{6}$. All elements observed in the sample and the gold foil supports are indicated. (b) and (c) display the high resolution Co 2$p$ and Mn 2$p$ regions and the corresponding peak components fitting using a Shirley background and two Lorentzian-Gaussian (30) peaks components.}
\label{FigXPS}
\end{figure}

In the Fig. \ref{FigXPS}(a) the XPS survey spectra for La$_{1.5}$Ca$_{0.5}$CoMnO$_{6}$ is shown and all elements found are indicated, including the gold foil support signal. Due to the polycrystalline nature of the material the carbon signal persist on the sample's surface and any attempt to gentle Argon sputter the surface results on species chemical reduction. Anyhow, the XPS still serves as a powerful tool to investigate the electronic structure of the cobalt and manganese cations in the compound. Based on the relatively large electron mean free path for the Co 2$p$ and Mn 2$p$ photoelectrons (larger than 1.5 nm) the spectra shown reveals chemical information from buried oxide layers \cite{Raveau,Petitto,Elp,Jimenez,Dudric,Koethe}. At the present conditions, although a detailed analysis of the cobalt and manganese 2$p$ regions are difficult, one can yet address the Co and Mn chemical state based on the spectral features.

It has been largely reported in the literature that the expected binding energies for Co$^{3+}$ and Co$^{2+}$ are located at 779.5 and 780.5 eV, respectively \cite{Hassel}. Noteworthy, depending on the cation symmetry one may observe intense satellite features related to Co$^{2+}$  at octahedral sites, in contrast to Co$^{3+}$ at octahedral or Co$^{2+}$ at tetrahedral sites \cite{Vaz}. The Co 2$p$ spectra shown in figure \ref{FigXPS}(b) presents prominent peaks at 779.7 and 795.0 eV related to the Co 2$p_{3/2}$ and 2$p_{1/2}$ spin-orbit components, respectively. Remarkable however is the lack of satellite features which in various compounds is a clear signature of Co$^{2+}$ in octahedral sites, most probably at high-spin configuration \cite{Brundle}.

The detailed analysis of the Co 2$p$ region is quite useful to determine the electronic structure of the cation. This analysis is however far from trivial, and the main peaks are known to be composed by multiplet splitting as described by Gupta and Sen \cite{Gupta}, and extensively discussed in several recent studies \cite{Biesinger}. Nevertheless, we have considered that the intensity of each spin-orbit component can be fitted using two curves related to the different Co cations and respective satellites since Co$^{2+}$ and Co$^{3+}$ are known to display well distinguish binding energies at about 780.1 eV and 779.6 eV. The spectral analysis for La$_{1.5}$Ca$_{0.5}$CoMnO$_{6}$ employing the two fitting components results on the assignment of Co$^{2+}$ and Co$^{3+}$ species located at binding energies and FWHM of 780.1 (2.84)  and 779.3 (1.82) eV, respectively. The fitted components can hardly determine the absolute amounts but serves to indicate the ratio of Co$^{3+}$/Co$^{2+}$ cations in the compound, about 1.3.

The Mn 2$p$ core level spectrum is shown in Fig. \ref{FigXPS}(c). The splitting between the 2$p_{3/2}$ and 2$p_{1/2}$ levels is about 11.5 eV, and are peaking at 641.5 eV and 653.0 eV, respectively. Once again, the relative concentration of Mn cations oxidation state is difficult to assign but the spectral features observed are indicative of Mn at either 3+ or 4+ valence states \cite{Oku,Castro}. The main peaks are fitted with two components indicating the relevant peak features. The presence of a shoulder on the high energy side and the relatively broad and symmetric peak is allusive of Mn in manganite \cite{Ardizzone}. We believe the Mn species are probably in the Mn$^{4+}$ oxidation state since no evidence of Mn$^{2+}$ or Mn$^{3+}$, such as satellite features, are observed and the typical peak asymmetry of Mn$^{4+}$ is only suppressed by the hydroxylation of the surface during sample cleaning \cite{Nesbitt,Ilton}. The XPS results support our XRD and magnetometry findings on the valence state of the Co and Mn cations in the compound, which can be interpreted in terms of Co$^{2+}$/Co$^{3+}$ mixed valence state (see below).

\subsection{AC and DC Magnetization}

Fig. \ref{FigMxT}(a) shows the ZFC and FC $T$ dependent magnetic susceptibility ($\chi$) for La$_{1.5}$Ca$_{0.5}$CoMnO$_{6}$, obtained at $H=100$ Oe. For $T$ $\geq$ 200 K, the data could be well fitted to the Curie-Weiss law [inset of Fig. \ref{FigMxT}(a)]. The effective magnetic moment obtained from the fit is $\mu_{eff}$ = 6.9 $\mu_{B}$/f.u., which can be compared to the value estimated from the usual equation for systems with two or more different magnetic ions \cite{Meu,Niebieskikwiat}
\begin{equation}
\mu = \sqrt{{\mu_1}^2 + {\mu_2}^2 + {\mu_3}^2 +...}. \
\end{equation}
Assuming $\sim$50\%/50\% proportion of high spin (HS) Co$^{2+}$/Co$^{3+}$ and complete quenching of orbital angular moments for the TM ions located at an octahedral environment ($\mu_{Co^{2+}}$ = $\mu_{Mn^{4+}}$ = 3.9 $\mu_{B}$,  and $\mu_{Co^{3+}}$ = 4.9 $\mu_{B}$) \cite{Ashcroft}, the estimated magnetic moment is $\mu$ = 5.9 $\mu_{B}$/f.u., which is smaller than the experimental value and may indicate the presence of Mn$^{3+}$ ($\mu_{Mn^{3+}}$ = 4.9 $\mu_{B}$). On the other hand, some orbital contribution for the moments is usually observed on Co- and Mn-based perovskites \cite{Raveau}, which must also be considered. Using the L+S moments for Mn$^{4+}$ (4 $\mu_{B}$) and HS Co$^{2+}$ and Co$^{3+}$ (5.2 and 5.5 $\mu_{B}$, respectively) \cite{Carlin} on Eq. 1 yields $\mu$ = 6.7 $\mu_{B}$/f.u., which is very close to the experimentally observed value.

\begin{figure}
\begin{center}
\includegraphics[width=0.5 \textwidth]{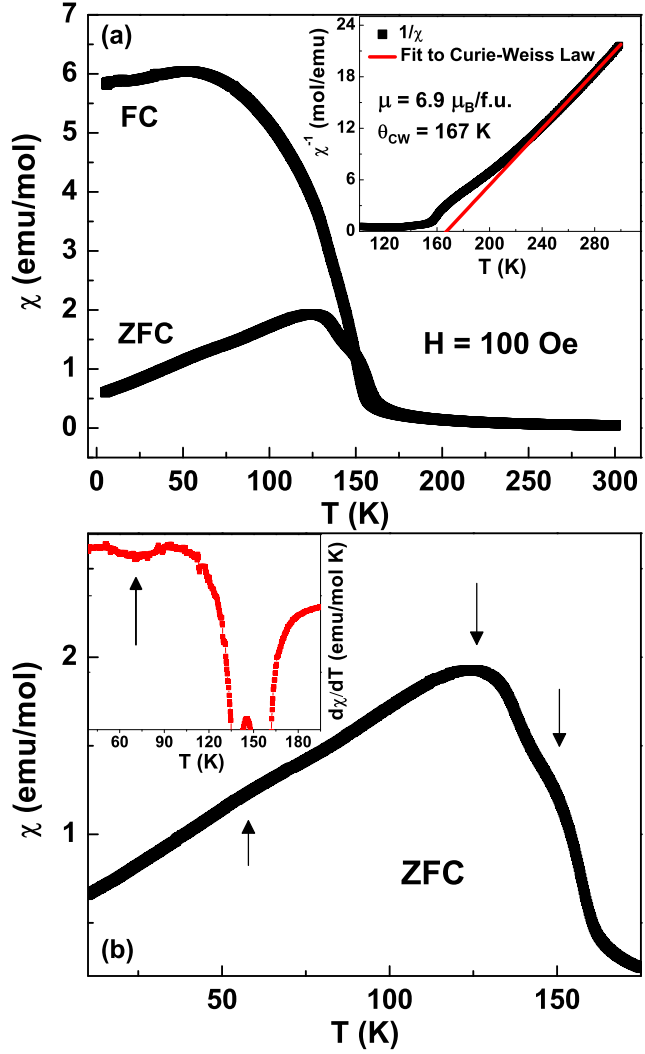}
\end{center}
\caption{(a) ZFC and FC $T$ dependent magnetic susceptibility. The inset shows the inverse of magnetic susceptibility and its fit to the Curie-Weiss law. (b) Magnified view of the ZFC curve at low-$T$, and the inset shows its derivative.}
\label{FigMxT}
\end{figure}

From the fitting of $\chi^{-1}$ it was also obtained the Curie-Weiss temperature, $\theta_{CW}\sim170$ K. The positive value indicates that the FM coupling dominates the magnetic interactions on the system. But a magnified view of the ZFC curve [Fig. \ref{FigMxT}(b)] reveals two anomalies above 100 K, which could be associated with the Co mixed valence state. The Co$^{2+}$--O--Mn$^{4+}$ superexchange interaction is predicted to be FM \cite{Goodenough}. The 25\% of Ca$^{2+}$ to La$^{3+}$ substitution on La$_{2}$CoMnO$_{6}$ induces the emergence of Co$^{3+}$, and the Co$^{3+}$--O--Mn$^{4+}$ coupling is expected to be AFM. Hence the second anomaly observed on Fig. \ref{FigMxT}(b) could be related to the emergent AFM ordering. It can also be noted a slight change in the slope of the curve at $T\sim55$ K. It becomes more evident when the derivative of the curve is carried, as indicated on the inset of Fig. \ref{FigMxT}(b). The presence of competing magnetic phases, together with the disorder that intrinsically accompanies the DP compounds, are understood as key ingredients to induce glassy states. As will be discussed next, this inflection point is related to the emergence of a CG state at low-$T$.

\begin{figure}
\begin{center}
\includegraphics[width=0.5 \textwidth]{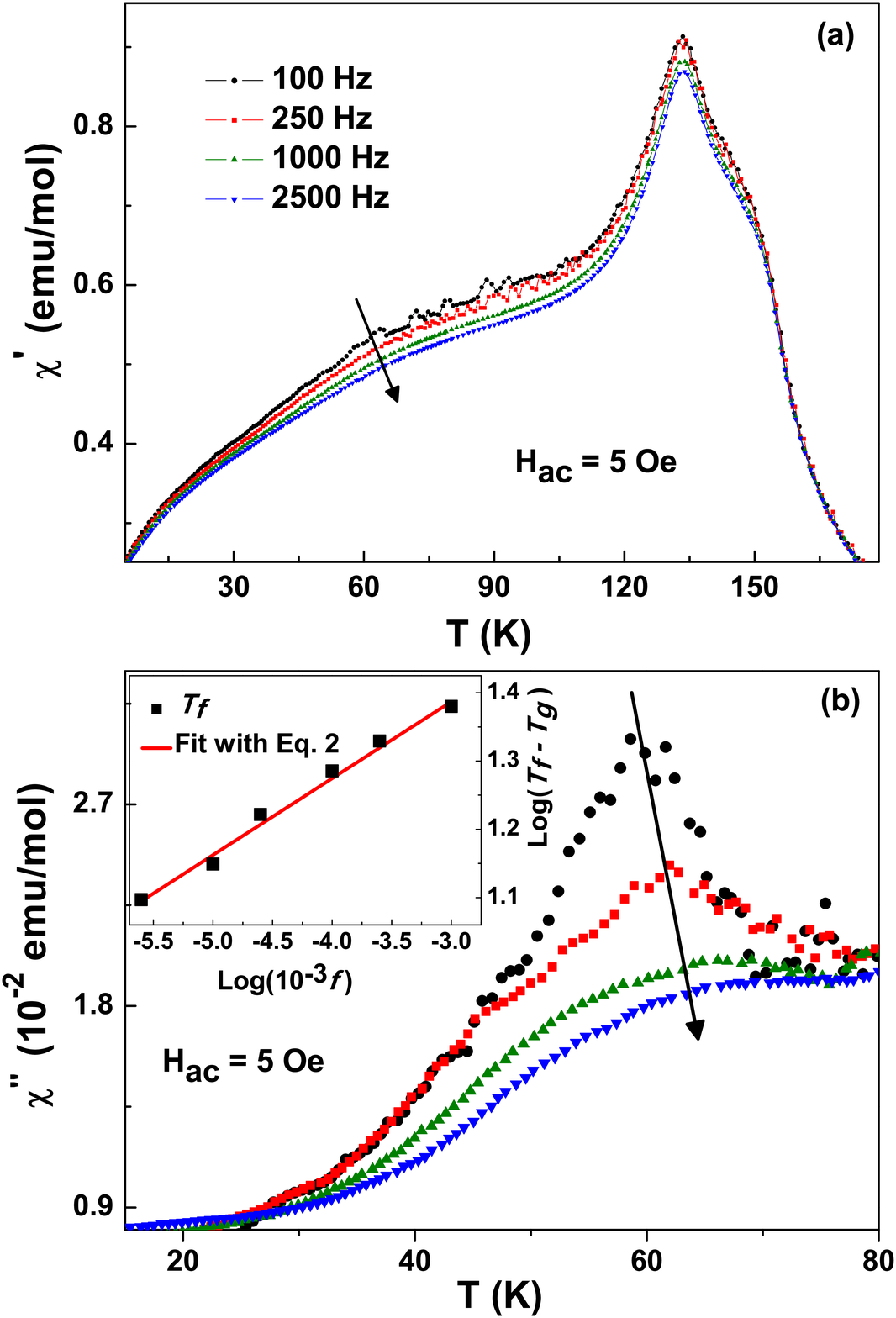}
\end{center}
\caption{(a) $\chi$'$_{ac}$ and (b) $\chi$''$_{ac}$ as a function of $T$ for some selected frequencies. The curves were obtained with $H_{ac}$=5 Oe oscilating field.  The inset shows $T_{f}$ as a function of $f$, where the solid line represents the best fit to the data using Eq. \ref{Eq2}.}
\label{FigChiAC}
\end{figure}

In order to verify the presence of a SG-like phase, which is believed to play an important rule on the ZEB effect, we have measured AC magnetic susceptibility as a function of $T$ for six frequencies ($f$) in the 25-10000 Hz range. Fig. \ref{FigChiAC} shows the real ($\chi$') and imaginary ($\chi$'') parts of the ac susceptibility for some selected frequencies.  Fig. \ref{FigChiAC}(a) shows variations at low-$T$ in the $\chi$' curves for different $f$, indicative of glassy behavior. For the $\chi$'' curves on Fig. \ref{FigChiAC}(b) one can see a peak ($T_{f}$ - freezing $T$) that decreases in magnitude and shifts to higher $T$ as $f$ increases. This is characteristic of SG-like systems \cite{Mydosh}. The evolution of $T_{f}$ with $f$ turns out to be well described by the power law equation of the dynamical scaling theory
\begin{equation}
\frac{\tau}{\tau_{0}}=\left[\frac{(T_{f} - T_{g})}{T_{g}}\right]^{-z\nu} \label{Eq2}
\end{equation}
where $\tau$ is the relaxation time corresponding to the measured $f$, $\tau_{0}$ is the characteristic relaxation time of spin flip, $T_{g}$ is the transition temperature ($T_{f}$ as $f$ approaches zero) and $z\nu$ denotes the critical exponent \cite{Mydosh}. The solid line on inset of Fig. \ref{FigChiAC}(b) represents the fit to Eq. \ref{Eq2}, yielding $T_{g}\simeq46.7$ K, $\tau_{0}\simeq3.1\times10^{-7}$ s and $z\nu\simeq8.9$, being these values typical of CG systems \cite{Murthy2,Mydosh}.

Usually, a formula computing the shift in $T_{f}$ between the two outermost frequencies is used to classify the material as SG, CG or superparamagnet
\begin{equation}
\delta T_{f}=\frac{\bigtriangleup T_{f}}{T_{f} \bigtriangleup log f}. \label{Eq3}
\end{equation}
For CG systems, $\delta$ usually lies in the range $0.01\lesssim\delta T_{f}\lesssim0.1$ \cite{Meu,Mydosh, Anand2,Malinowski}, and for La$_{1.5}$Ca$_{0.5}$CoMnO$_{6}$ we found $\delta\simeq$0.07. This CG phase, together with the conventional magnetic transitions observed at higher $T$, confirms the RSG state on this material.

\subsection{Exchange bias}

The $M(H)$ curves were measured at various $T$. In order to prevent the presence of remanent magnetization, it was followed a careful protocol of sending the sample to the paramagnetic state and demagnetizing the system in the oscillating mode between one measurement to another. This protocol is particularly important for the ZFC $M(H)$ measurements, since any trapped current in the magnet would prevent the correct observation of the ZEB effect. Moreover, to get a reliable comparison between the results obtained at different $T$, all measurements were carried with the same experimental protocol, i.e., the same scan length, maximum moment range, frequency, $T$-rate on the ZFC procedure etc.

\begin{figure}
\begin{center}
\includegraphics[width=0.5 \textwidth]{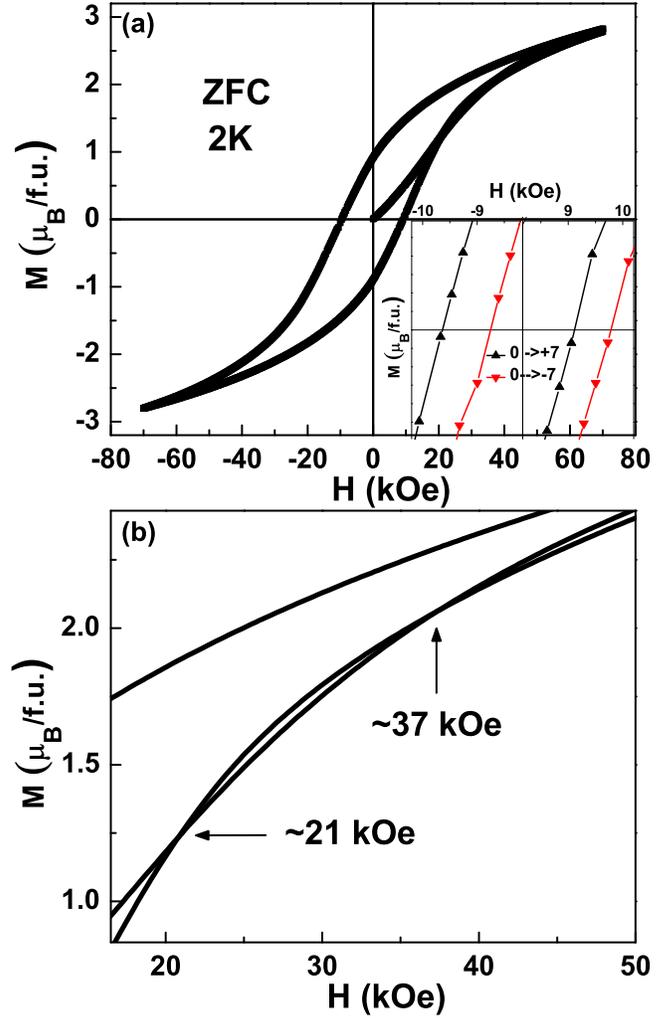}
\end{center}
\caption{(a) ZFC $M(H)$ loop at 2 K. The inset shows a magnified view of the loops measured in the modes $0\rightarrow7$T$\rightarrow-7$T$\rightarrow7$T and $0\rightarrow-7$T$\rightarrow7$T$\rightarrow-7$T. (b) Magnified view of the intersections between the virgin curve and the second branch of the loop.}
\label{FigMxH}
\end{figure}

Fig. \ref{FigMxH}(a) shows the curve measured at 2 K, obtained after cooling the sample in ZFC mode. It is a closed loop that does not saturate even for the maximum applied field $H_{max}$ = 70 kOe, which can be explained by the presence of AFM and CG phases. The large coercivity masks the curves asymmetry for the complete loop, but a magnified view on inset of Fig. \ref{FigMxH}(a) reveals the negative shift of the $M(H)$ curve (i.e., the shift direction is opposite to the positive field initially applied in the virgin curve). To verify this effect we have measured $M(H)$ with the initial $H$ in the opposite  (negative) direction. As can be observed on inset of Fig. \ref{FigMxH}(a), the curve exhibits a positive shift. The shift in the opposite direction is an expected behavior of a EB system, hence a clear evidence that this result is intrinsic of the material.

The ZEB effect is believed to be related to the field induced reconfiguration of the spins at the magnetic interfaces that occurs during the initial magnetization of the system \cite{Wang,Nayak,Maity,Murthy,Meu}. Thus in the $M(H)$ curve of a ZEB material the virgin curve usually lies outside the main loop for a certain $H$ interval, and for some critical $H$ the curves intersect \cite{Wang,Nayak,Murthy}. This can be seen for La$_{1.5}$Ca$_{0.5}$CoMnO$_{6}$ on Fig. \ref{FigMxH}(b), where the virgin curve and the second branch of the hysteresis loop intercepts at $H$ $\sim$ 21 kOe. But differently than for other ZEB materials, at $H$ $\sim$ 37 kOe the two curves cross again, indicating a second field-induced reconfiguration. This suggests the possible presence of two contributions to the ZEB effect, leading to a decrease of the final magnetization which is probably related to the smaller ZEB effect observed for La$_{1.5}$Ca$_{0.5}$CoMnO$_{6}$ in comparison to La$_{1.5}$Sr$_{0.5}$CoMnO$_{6}$.

\begin{figure}
\begin{center}
\includegraphics[width=0.5 \textwidth]{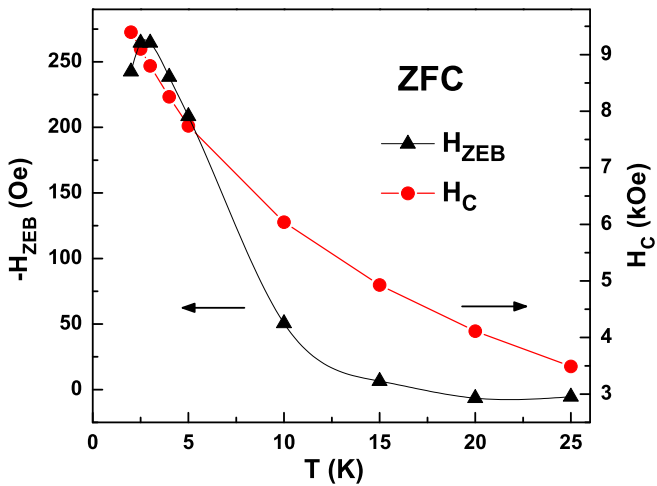}
\end{center}
\caption{-$H_{ZEB}$ and $H_{C}$ evolution with $T$. The lines are guides for the eye.}
\label{FigZEB}
\end{figure}

The EB field is here defined as $H_{EB}=(H^{+}+H^{-})/2$, where $H^{+}$ and $H^{-}$ represent the positive and negative coercive fields,  respectively. The effective coercive field is $H_{C}=(H^{+}-H^{-})/2$. Fig. \ref{FigZEB} shows that a negative ZEB effect is observed for La$_{1.5}$Ca$_{0.5}$CoMnO$_{6}$ for $T<15$ K, in resemblance to its analogue La$_{1.5}$Sr$_{0.5}$CoMnO$_{6}$ \cite{Murthy}. But here the magnitude of the ZFC EB field, $H_{ZEB}$, is much smaller than for the Sr-based compound, which presents the largest ZEB effect reported so far. At $T$ = 2 K one have $H_{ZEB}\sim$ 250 Oe for Ca-based sample, while for Sr-based compound $H_{ZEB}\sim$ 6.5 kOe.  Also interesting to note is the fact that initially $H_{ZEB}$ increases with $T$. This result is similar to that observed for the perovskite-based ZEB nanocomposite BiFeO$_{3}$-Bi$_{2}$Fe$_{4}$O$_{9}$ \cite{Maity}. Possibly, the small gain of thermal energy at low-$T$ enables the rotation of some AFM/CG spins to the field direction, leading to the pinning of these spins after the removal of magnetic field.

\begin{figure}
\begin{center}
\includegraphics[width=0.5 \textwidth]{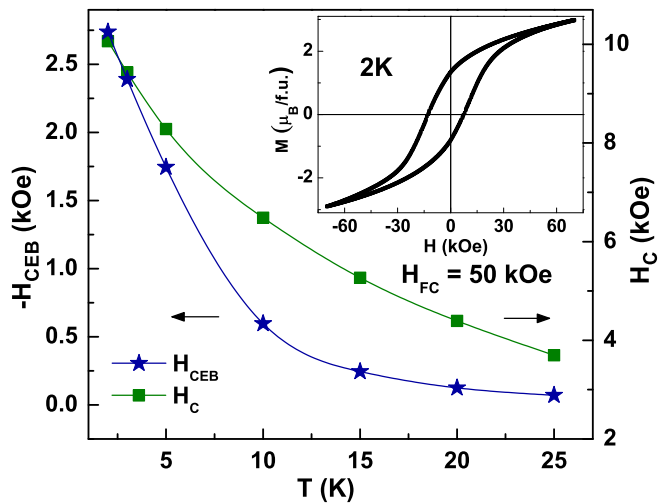}
\end{center}
\caption{-$H_{CEB}$ and $H_{C}$ evolution with $T$, after cooling the sample with $H_{FC}=50$ kOe. The lines are guides for the eye. The inset shows the FC $M(H)$ loop at 2 K.}
\label{FigCEB}
\end{figure}

In order to further investigate the EB effect on La$_{1.5}$Ca$_{0.5}$CoMnO$_{6}$, we have also carried $M(H)$ curves after cooling the sample in the presence of a field $H_{FC}=50$ kOe. The inset of Fig. \ref{FigCEB}(a) shows the curve obtained at 2 K, for which was observed an EB field $H_{CEB}$ = 2740 Oe. This signifies a great enhancement in comparison to the ZFC measurement, and resembles results observed for related compounds as La$_{1.5}$Sr$_{0.5}$CoMnO$_{6}$ and La$_{1.5}$Ca$_{0.5}$CoIrO$_{6}$, for which $H_{FC}$ induces the pinning of the spins already at high $T$, enhancing the EB effect \cite{Murthy,Meu}.

In EB systems, $H_{EB}$ generally depends on the number of $M(H)$ measurements, a property often called training effect \cite{Nogues}. The possible application of an EB material for spintronic devices requires it to be fairly stable under consecutive $M(H)$ cycles. Since La$_{1.5}$Ca$_{0.5}$CoMnO$_{6}$ exhibits both ZEB and CEB effects, we have investigated the dynamics of the spin structure by measuring consecutive $M(H)$ loops at 2 K in both ZFC and FC cases.

\begin{figure}
\begin{center}
\includegraphics[width=0.5 \textwidth]{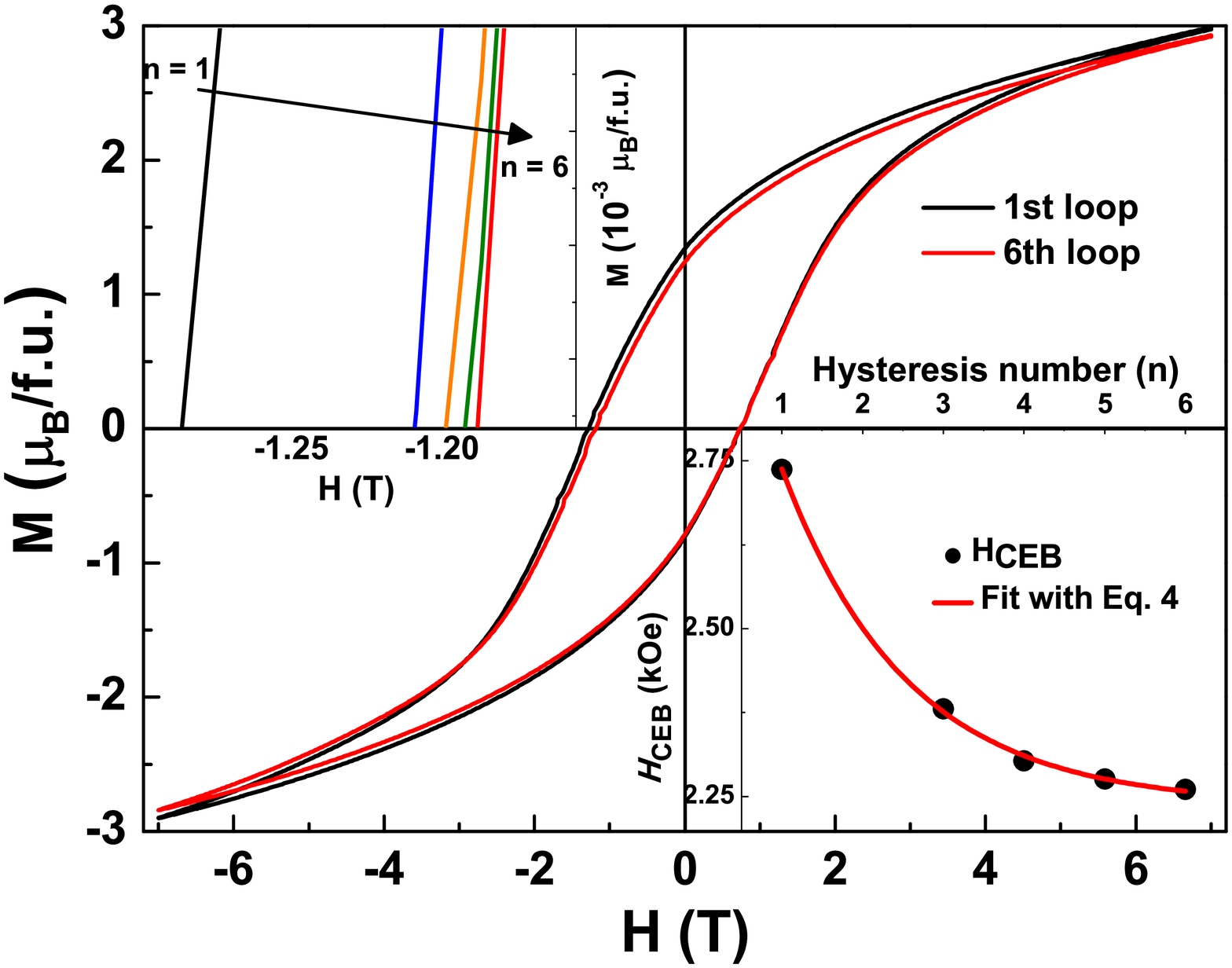}
\end{center}
\caption{First and sixth loops obtained at 2 K, after cooling the system with $H_{FC}$ = 50 kOe. The upper inset shows a magnified view of $H^{-}$ for six consecutive curves. The bottom inset shows -$H_{CEB}$ as a function of the hystereis number ($n$). The solid line represents the fitting of experimental data to Eq. \ref{EqTE2}.}
\label{FigTE}
\end{figure}

For the CEB mode, the consecutive loops were carried after cooling the system with $H_{FC}$ = 50 kOe. Fig. \ref{FigTE} shows the first and sixth loops, where can be noted that the differences in the $H_{CEB}$ fields are mainly due to variations in $H^{-}$. On the upper inset of Fig. \ref{FigTE} it becames clearer that the magnitude of $H^{-}$ monotonically decreases with the number of consecutive cycles ($n$), indicating spin rearrangement at the interfaces. The bottom inset of Fig. \ref{FigTE} shows the evolution of $H_{CEB}$ with $n$. This curve can be fit to a model considering two contributions at the magnetic interfaces, the uncompensated rotatable spins and also the frozen spins \cite{Meu,Mishra}
\begin{equation}
|{H}^n_{CEB}| = |H^{\infty}_{CEB}| + A_{f}e^{(-n/P_{f})} + A_{r}e^{(-n/P_{r})}, \label{EqTE2}
\end{equation}
where $f$ and $r$ denote the frozen and rotatable spin components respectively. The fitting with  Eq. \ref{EqTE2} yields $H^{\infty}_{CEB}\sim2238$ Oe, $A_{f}\sim951$ Oe, $P_{f}\sim1.56$, $A_{r}\sim100$ Oe, $P_{r}\sim0.04$. The precise fitting obtained when the glassy phase is taken into account, together with the fact that all ZEB materials reported so far presents a SG-like phase concomitantly to another magnetic phase, indicate the importance of the glassy phase to the observation of EB effect on this and related PD compounds.

\begin{figure}
\begin{center}
\includegraphics[width=0.5 \textwidth]{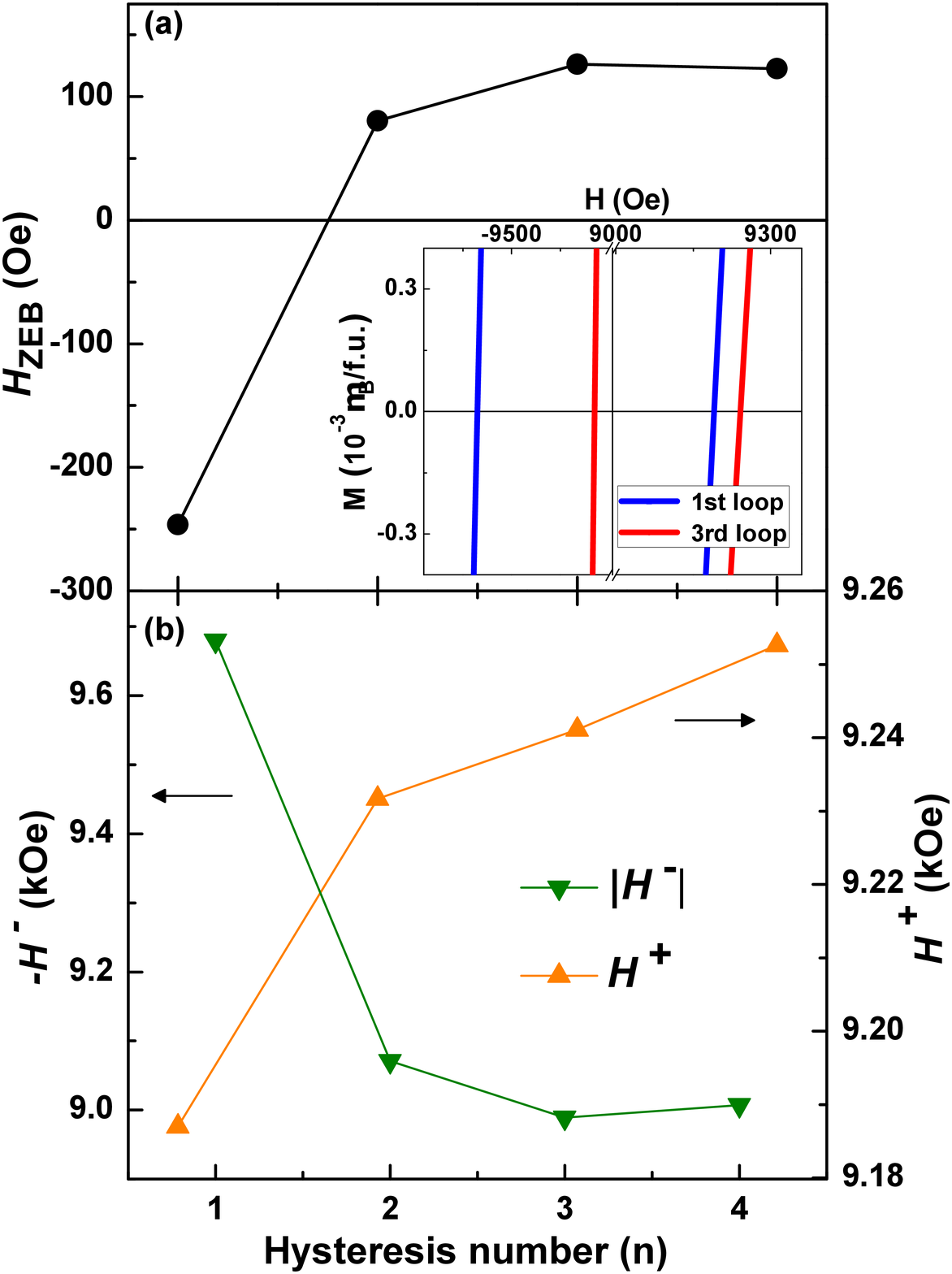}
\end{center}
\caption{(a) $H_{ZEB}$ as a function of the hysteresis number ($n$). The inset shows a magnified view of the first and third curves close to the $M=0$ axis. (b) $H^{+}$ and -$H^{-}$ as a function of $n$, obtained after the ZFC protocol. The lines are guides for the eye.}
\label{FigZFCTE}
\end{figure}

For the ZFC mode, an unprecedented behavior was observed. As Fig. \ref{FigZFCTE}(a) shows, from the first to the second loop, $H_{ZEB}$ inverts its sign from negative to positive. This result indicates that the ZFC procedure does not ensure a good stability in the coupling between the spins responsible for the EB effect. From the second loop to subsequent measures, the system maintains a roughly constant positive value. A closer inspection on the positive and negative coercive fields shows that the absolute value of $H^{-}$ decreases for consecutive loops, while $H^{+}$ increases [Fig. \ref{FigZFCTE}(b)].

Although this is the first report of sign inversion on the EB field after the ZFC procedure, this behavior was already observed for CEB systems. For NiFe/IrMn bilayer, the training-induced inversion of $H_{CEB}$ signal was discussed in terms of the appearance of metastable magnetic disorder at the magnetically frustrated interface during the hysteresis loop measurements\cite{Mishra}. Similarly to the SG model conjectured for EB systems in Refs. \cite{Mishra,Radu}, our results can be understood in terms of two contributions to the EB effect, the rotatable and frozen spins at the AFM/FM/CG interfaces. For successive $M(H)$ measurements, the rearrangement of the rotatable AFM/FM spins at the interface would lead to the expected decrease on the magnitude of $H^{-}$ observed on Fig. \ref{FigZFCTE}(b). The second contribution would arise from the CG/FM interface. Some spins at this interface will prefer to align antiparallel to the direction of the FM phase, defined by the positive field firstly applied in the first cycle. After training, these spins may rotate irreversibly due to the magnetization reversal of the FM phase, leading to the increase of $H^{+}$ and the consequent inversion of $H_{ZEB}$ sign.

\section{CONCLUSIONS}

In summary, La$_{1.5}$Ca$_{0.5}$CoMnO$_{6}$ is a RSG material for which there are two anomalies on the magnetization as a function of $T$ curve at $T\simeq158$ K and $T\simeq138$ K, probably related to Co$^{2+}$--O--Mn$^{4+}$ FM coupling and Co$^{3+}$--O--Mn$^{4+}$ AFM coupling. At $T_{g}\simeq46.7$ K, a CG phase emerges. Similarly to the analogous La$_{1.5}$Sr$_{0.5}$CoMnO$_{6}$ RSG compound, La$_{1.5}$Ca$_{0.5}$CoMnO$_{6}$ exhibit a non-negligible ZEB effect at low-$T$. But the negative shift of the $M(H)$ curve observed for the Ca-based material is substantially smaller than for Sr-based one. We believe that the lower symmetry of the Co--O--Mn coupling for the Ca-doped sample and the particular behavior of its initial magnetization observed at the $M(H)$ curve are responsible for such difference. When the system is cooled in the presence of an applied magnetic field, the EB effect is greatly enhanced. The fitting of the training effect observed after the FC protocol indicates the importance of the glassy phase to the EB observed on this sample. For the ZFC procedure, $H_{ZEB}$ inverts its sign from negative to positive from the first to the second loop, which may be related to the unstable glassy spins at the interfaces. The presence of SG-like phase concomitantly to other magnetic phases seems to be a necessary condition to the observation of ZEB. This result should be taken into account when designing materials for application of the ZEB effect.

\begin{acknowledgments}
This work was supported by CNPq, CAPES, FAPESP and FAPERJ (Brazil). L. Bufai\c{c}al thanks prof. S. A. Le\~{a}o for the support. F. Stavale thanks the Surface and Nanostructures Multiuser Lab and the Lab of Biomaterials at CBPF.
\end{acknowledgments}

\end{document}